\documentclass[prb,twocolumn,superscriptaddress,showpacs,floatfix]{revtex4}
\usepackage{graphics}
\usepackage{graphicx}
\usepackage{amssymb}
\usepackage{color}

\begin{document}

\title{Impact of the valley degree of freedom on the control of donor electrons near a
Si/SiO$_2$ interface}
\author{A. Baena}
\affiliation{Instituto de Ciencia de Materiales de Madrid, ICMM-CSIC,
Cantoblanco, E-28049 Madrid, Spain}
\author{A. L. Saraiva}
\affiliation{Instituto de Fisica, Universidade Federal do Rio de Janeiro,
Caixa Postal 68528, 21941-972 Rio de Janeiro, Brazil}
\author{Belita Koiller}
\affiliation{Instituto de Fisica, Universidade Federal do Rio de Janeiro,
Caixa Postal 68528, 21941-972 Rio de Janeiro, Brazil}
\author{M.J. Calder\'on}
\affiliation{Instituto de Ciencia de Materiales de Madrid, ICMM-CSIC,
Cantoblanco, E-28049 Madrid, Spain}
\date{\today}

\begin{abstract}
We analyze the valley composition of one electron bound to a shallow donor
close to a Si/barrier interface as a function of an applied electric
field. A full six-valley effective mass model Hamiltonian is adopted.
For low fields, the electron ground state is essentially confined at the donor. At high fields the ground state is such that  the  electron is drawn to the interface, leaving the donor practically ionized. Valley splitting at the interface occurs
due to the valley-orbit coupling, $V_{\rm vo}^I = |V_{\rm vo}^I|e^{i\theta}$. At
intermediate electric fields, close to a  characteristic shuttling
field, the electron states may constitute hybridized states
with valley compositions different from the donor and the interface
ground states.
The full spectrum of energy levels shows crossings and anti-crossings
as the field varies. The degree of level repulsion, thus the width of
the anti-crossing gap, depends on the relative valley compositions,
which vary with $|V_{\rm vo}^I|$, $\theta$ and the interface-donor distance. We focus on the valley configurations of the states involved in
the donor-interface tunneling process, given by the anti-crossing of
the three lowest eigenstates. A sequence of two
anti-crossings takes place and the complex phase $\theta$ affects
the symmetries of the eigenstates and level anti-crossing gaps.
We discuss the implications of our results on the practical
manipulation of donor electrons in Si nanostructures.

\end{abstract}
\maketitle
\section{Introduction}
\label{sec:intro}

The search for a functional quantum computer (QC) started in the
mid-nineties, and by the year 2000 many systems had been considered as
candidates for its physical implementation.\cite{nielsenchuang} Among them,
the 1998 proposal for a Si-based QC by Kane~\cite{kane98}  raised special
interest due to objective and relevant factors favoring Si, such as the
accumulated know-how in processing Si for advanced device applications, the
relatively long spin coherence times and the possibility of isotopic
purification processing, further increasing coherence times.\cite{sarma04}

On the other hand, the conduction electrons in Si are not
in a well defined single Bloch state. Instead, the Si conduction
band is six-fold degenerate, with minima (valleys) along the $x$, $y$
and $z$ crystallographic directions. This imposes
limitations to the spin manipulation and coherence.\cite{Divincenzo}

It was recently proposed to encode quantum information directly into the
valley degree of freedom, converting the spurious valley Hilbert subspace
into a useful ingredient for a QC.\cite{dimi2012} Naturally, this raises fundamental
questions, such as how to promote controlled inter-valley transitions, to
what extent valley degeneracy can be lifted, and how sensitive such operations
are to fabrication-related parameters. The valley degree of freedom also affects
transport properties in Si nanostructures: valley degeneracy  has been recently shown to produce
a valley Kondo effect in a singly doped Si fin field effect transistor.\cite{tettamanzi2012}

We study here
the valley degree of freedom for one electron bound to a donor
--- more specifically substitutional P in Si --- tunnel-coupled to a (001)
Si/SiO$_2$ interface at a distance $d$ from the donor. The barrier material
is taken to be SiO$_2$ for definiteness, but it could in principle be any high
quality interface, such as Si/SiGe. The evolution of the inequivalent valley
contributions is obtained by mapping the low-lying manifold (following in
more detail the three lowest energy states) as an electric field pulls the
electron away from its ``hydrogenic'' configuration at the donor site towards
the state at the interface.

A ground state  electron confined in the direction perpendicular to the interface in the
triangular potential formed by the barrier and the electric field [see Fig.~\ref{Esquema}(a)],
still remains bound to the donor core potential, leading to localization in the
in-plane direction. \cite{calderon-longPRB07,calderonPRL06} For low fields, the electron in the ground state
is essentially confined at the donor, where the lowest energy manifold is split by the tetrahedral
crystal field environment into states with distinct contributions from
the six valleys. In particular, the ground state at the donor is a non-degenerate symmetric combination of the six valleys.

At high enough fields, the donor is ionized, and its electron
is shuttled to the interface. At the interface, the valley levels split into a four-fold
degenerate excited state, consisting of $\pm x$ and $\pm y$ Bloch states,
and a lower manifold spanned by the $\pm z$ valleys. The two lowest levels
are only slightly separated due to the abrupt interface breaking the $z$ reflection
symmetry (generally less than $1$ meV separation~\cite{sham79,goswami2007,chutia2008,Andre2009,dzurak2011,Andre2011}).

These relatively simple and well understood valley compositions at interfaces
and isolated donors could lead to a variety of  compositions at
intermediate fields, as illustrated by the main panel in Fig.~\ref{Esquema}.
 The lines are symmetry-allowed paths connecting energy levels from the
low-field (right of the panel) to the high-field (left) regime: one may
anticipate a rich variety of behaviors and formation of hybrid
donor-interface states.  Each level crossing or anti-crossing as the field
varies may change the valley compositions of the involved states.

The Stark shifted spectrum of P donors in bulk Si (no barrier material) is discussed in Ref.~[\onlinecite{debernardi06}]. The effective mass approach is adopted and the envelope functions are expressed as a combination of atomic-like orbitals.   The same group considered later the Stark effect for P donors at the center of Si nanospheres embedded in a barrier material. \cite{debernardi2010} The quasi-spherical symmetry of the problem without field allows the assignement of hydrogenic quantum numbers to the interface states. Tight-binding calculations for systems more directly related to the present  geometry were reported in Ref.~[\onlinecite{rahman2009}]. In these references elaborate numerical procedures are adopted leading to accurate results. Here we get good qualitative agreement for the spectrum in comparison with these previous works, and we explore a complementary aspect - namely the valley-composition evolution of the low-lying states under an increasing external field. The same geometry has been previously studied in a two-valley model $\{(000001),(000010)\}$ where the valley-composition analysis is not accessible. \cite{calderonPRB08} 

We discuss here the evolution of the
valley quantum number as a function of the electric field for field values in the range  where the hybridization between donor and interface states takes place. We show that the phase of the interface valley-orbit coupling plays an important role in defining the gap amplitudes, affecting manipulation capabilities for the donor electrons. Our model sheds
light to the qualitative features of the level diagrams shown in
Refs.~[\onlinecite{debernardi06,rahman2009}],
and permits to predict the expected diagrams for various geometries
of the donor/barrier problem. Our aim is to get a clear description of the changes in symmetry and
valley composition of the states involved in the donor-interface electron shuttling problem. Our results may be useful for
donor electron valley manipulation via an applied field.

This paper is organized as follows. In Sec.~\ref{sec:model} the system is
described and a 6-valley effective mass model-Hamiltonian for the donor
electron under an applied electric field is explained. In
Sec.~\ref{sec:results}, our results for the electric field dependence of the
low-lying spectrum and of the valley composition of the three lowest
electronic eigenstates are presented. We conclude in
Sec.~\ref{sec:discussion} with a discussion of the implications of these
results on the practical manipulation of donor electrons in Si
nanostructures.

\section{Model}
\label{sec:model}
\begin{figure}
\includegraphics[clip,width=0.55 \textwidth]{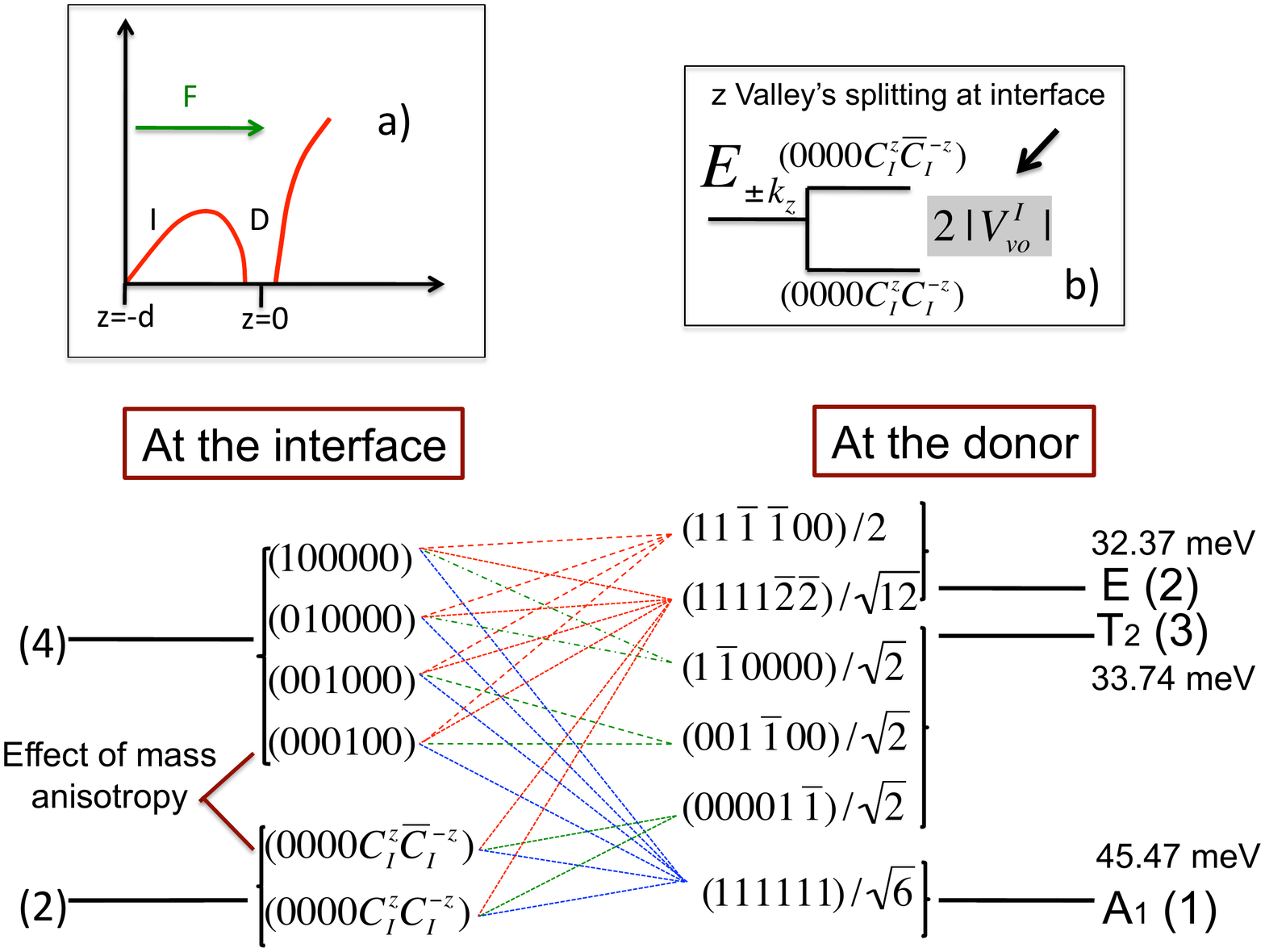}
\caption{(Color online) (a) Double well potential in the z-direction formed by the Coulombic
 donor potential plus the triangular interface/electric field potential.
  $d$ is the distance between the donor and interface.
  (Main panel) Symmetry of levels at the donor and at the interface.
  Every level is described by six coefficients corresponding to the six valleys
  of Si conduction band: $(x,-x, y, -y, z, -z)$.
  This defines the valley composition of each state. At the interface, the mass
  anisotropy breaks the valleys degeneracy in a doublet ($z,-z$) and a quadruplet ($x,-x,y,-y$).
  The doublet degeneracy is lifted due to the valley orbit coupling $(V^{I}_{\rm vo}=|V_{\rm vo}^I| e^{i \theta})$ arising in a sharp
  $(001)$ interface,~\cite{Andre2009,sham79,Andre2011} as shown in (b). $C_{I}^z$ and $C_{I}^{-z}$ are defined in Eq.~(\ref{eq:inter-full}), and $\bar C_I^{-z}=-C_I^{-z}$.
  At an isolated donor, the valley orbit coupling leaves a non-degenerate ground
  state with $A_1$ symmetry, well separated (splitting $\sim 12$ meV) from the other five levels. \cite{kohn55,ning71-I,koiller02PRL} The binding energies given on the right of each level are experimental values for bulk P donors.\cite{ning71-I} The lines join the valley compositions at donor and interface that are connected by symmetry.  
}
\label{Esquema}
\end{figure}

We consider a single electron bound to a substitutional P donor at $z=0$ near
a Si $(001)$ / SiO$_2$  interface at $ z=-d$ and under an applied uniform
electric field perpendicular to the interface, $\vec F = F \hat z$ pointing
from the barrier to Si, thus pushing the electron away from the donor and
towards the interface (see Fig.~\ref{Esquema}(a)). In effective ``atomic''
units for Si, $Ry^*=m_{\perp} e^4/2 \hbar^2 \epsilon_{\rm Si}^2=19.98$ meV
and $a^*=\hbar^2 \epsilon_{\rm Si}/m_{\perp} e^2=3.157$ nm, the Hamiltonian
is written as \cite{macmillen84,calderonPRL06}
\begin{equation}
H=-\frac{\partial^{2}}{\partial{x^2}}-\frac{\partial^{2}}{\partial{y^2}}-\gamma\frac{\partial^{2}}{\partial{z^2}}-\frac{2}{r}+keFz+H_{\rm vo}
\label{eq:hamil}
\end{equation}

with $\gamma=m_{\perp}/m_{\|}$ as the ratio between the transverse ($m_{\perp}=0.191m$) and longitudinal ($m_{\|}=0.916m$) effective masses, $\epsilon_{Si}=11.4$ , $k=3.89 \cdot
10^{-7}\epsilon_{Si}^{3}(m/m_{\perp})$cm/kV, and the electric field $F$ is
given in kV/cm. In Eq.~(\ref{eq:hamil}), the kinetic energy is
$(-\frac{\partial^{2}}{\partial{x^2}}-\frac{\partial^{2}}{\partial{y^2}}-\gamma\frac{\partial^{2}}{\partial{z^2}})$,
the next two terms are the donor Coulomb potential and the electric-field
linear potential, respectively. The last term describes the valley-orbit
effects, namely the coupling between different valleys due to the singular
nature of both the donor (D) and the interface (I) potentials, as described
below.

As shown in Fig.\ref{Esquema}(a), the system can be modeled by the combination of two potential wells: one that binds the electron to the donor (at low fields) and another binding the electron at the interface (at high fields).\cite{calderonPRL06} Note that the interface potential includes not only the electric field perpendicular to the interface but also the Coulomb attraction to the donor at a distance $d$ which confines the electron in the $xy$-plane even when it is at the interface.\cite{calderon-longPRB07} We simplify the calculation by initially computing the variational ground state wave-functions at the donor and at the interface, which define the envelopes in the basis for the full six valley problem.

The conduction band of Si has six degenerate minima (valleys) in the $\langle
100 \rangle$ directions at a distance $k_{0}=2 \pi \frac{0.85}{a_{\text Si}}$
from the $\Gamma$ point, where $a_{\text Si}=5.4$\AA~ is the lattice parameter
of Si.
In the simplest effective mass approximation, only the Bloch functions
at the positions of the conduction band minima are considered, and the ground
state of the electron at the donor is written \cite{kohn55}
\begin{equation}
\Psi_{D}=\sum_{\mu  = \pm x,\pm y,\pm z}C_D^{\mu} F_{D}^{\mu}(\mathbf{r})\phi_{\mu}(\mathbf{r}) \, ,
\label{eq:donor-full}
\end{equation}
where $F_{D}^{\mu}(\mathbf{r})=F_{D}^{-\mu}(\mathbf{r})$ are envelope functions and
$\phi_{\mu}(\mathbf{r})=u_{\mu} ({\bf{r}}) e^{i \bf{k}_{\mu} \cdot \bf{r}}$ are the six
Bloch eigenstates at the conduction band minima.
We take the variational donor envelope functions
$F_{D}^{\nu}(\mathbf{r})$, where $\nu = |\mu|$, centered at $r=0$, following the form introduced in
Ref.~[\onlinecite{kohn55}]
\begin{eqnarray}
F_{D}^{x}&=&N_D^{x}\, e^{-\sqrt{\frac{y^2+z^2}{a^2}+\frac{x^2}{b^2}}} \, , \\
F_{D}^{y}&=&N_D^{y}\, e^{-\sqrt{\frac{x^2+z^2}{a^2}+\frac{y^2}{b^2}}} \, ,\\
F_{D}^{z}&=&N_D^{z}\, e^{-\sqrt{\frac{x^2+y^2}{a^2}+\frac{z^2}{b^2}}} \, .
\label{eq:donor-envelopes}
\end{eqnarray}
These are normalized hydrogenic 1s envelopes, distorted due to the Si conduction band effective mass anisotropy, and $\{N_D^\nu\}$  are normalization factors. The effective Bohr radii $a$ and $b$ are variational parameters chosen to minimize the ground state energy. For the
distances used here ($d\gtrsim 2a^*\approx6$ nm), $a$ and $b$ coincide  with
Khon and Luttinger's variational parameters for a single
impurity in the bulk\cite{kohn55} $(d\rightarrow\infty)$, namely, $a= 2.365$ nm  and
$b=1.36$ nm.~\cite{calderon-longPRB07}

The six-fold degeneracy of the ground state is lifted at a substitutional
impurity because the translational symmetry of the host crystal is broken,
leading to intervalley scattering effects known as the valley-orbit
interaction.~\cite{baldereschi70,pantelides} This effect can be
accounted for phenomenologically, introducing a coupling between valleys in
perpendicular directions (e.g. $x,z$) $-\Delta_{c}$ and in parallel directions
(e.g. $z,-z$) $-\Delta_{c}(1+\delta_{c})$.\cite{koiller02PRB}
This splits the unperturbed
six-fold-degenerate donor electron ground state into a singlet ($A_1$
symmetry), a triplet ($T_2$ symmetry) and a doublet ($E$ symmetry), see Fig.~\ref{Esquema}. For P in Si, the
relative splittings between the different symmetry levels are reproduced taking
$\Delta_c=2.16$ meV and $\delta=-0.3$.\cite{koiller02PRB}

The degeneracy is also lifted near the (001) interface.\cite{ando82,kane00}
First, due to the mass anisotropy, the $z$ and $-z$ perpendicular valleys are
lower in energy than the ones parallel to the interface. The two lowest
energy states are combinations of the $z$ and $-z$ valleys, whose double
degeneracy is lifted due to valley-orbit coupling $V_{\rm vo}^I$ at an abrupt
interface. In general, $V_{\rm vo}^I$ is a complex quantity with an absolute value
proportional to the applied electric field\cite{sham79,Andre2009}  $|V_{\rm vo}^I|=\lambda F$, and dependent on the barrier height and
abruptness.\cite{Andre2009} The prefactor $\lambda$ has been estimated by
several authors.\cite{Andre2011}  We use initially the largest,
$\lambda=1.36$ \AA, as suggested in Ref. ~[\onlinecite{chutia2008}]. For instance,
for $F=50$ kV/cm, we have $|V_{\rm vo}^I|=0.68$ meV. We also consider the smaller value of  $\lambda= 0.215$\AA, estimated by Sham and Nakayama,~\cite{sham79} and discuss the qualitative changes that occur in the spectrum. The complex
phase of this valley-orbit coupling is also dependent on the interface
quality and has been estimated to be $\sim -1$ for an abrupt Si/SiO$_2$ interface.\cite{noteAndre2011ReIm}
$H_{\rm vo}$ in Eq.~(\ref{eq:hamil}) takes into account all these valley-orbit
interactions, both at the donor (D) and at the interface (I). Its form (in a
basis set defined next) is described in the Appendix.

Following Eq.~(\ref{eq:donor-full}), we take the lowest states at the interface as
\begin{equation}
\Psi_{I}=\sum_{\mu= \pm x,\pm y,\pm z}C_I^{\mu}F_{I}^{\mu}(\mathbf{r})\phi_{\mu}(\mathbf{r}) \, .
\label{eq:inter-full}
\end{equation}
The envelope functions $F_{I}^{\nu}(\bf{r})$, with $\nu = |\mu|$, are taken
in the variational form \cite{calderon-longPRB07}
\begin{eqnarray}
F_{I}^{x}&=&N_I^{x}(z+d)^2e^{-\alpha_{xy}(z+d)/2}e^{-(\beta_{1}x^2+\beta_{2}y^2)/2} \, , \\
F_{I}^{y}&=&N_I^{y}(z+d)^2e^{-\alpha_{xy}(z+d)/2}e^{-(\beta_{2}x^2+\beta_{1}y^2)/2} \, , \\
F_{I}^{z}&=&N_I^{z}(z+d)^2e^{-\alpha(z+d)/2}e^{-\beta^2\rho^2/2} \, ,
\end{eqnarray}
where the penetration into the barrier is considered to be negligibly small.
Here, $\rho^2=x^2+y^2$, and $\alpha$, $\alpha_{xy}$, $\beta$, $\beta_{1}$ and
$\beta_{2}$ are variational parameters: $1/\alpha$ and $1/\alpha_{xy}$ are
related to the width of the wave functions along the $z$-direction, and
depend on the value of the applied electric field; $\beta$, $\beta_{1}$ and
$\beta_{2}$ correspond to the confinement in the $xy$ plane, which is
controlled by the attractive potential of the donor, hence the ``$\beta$''
parameters depend on the distance $d$.\cite{calderon-longPRB07}

In Eqs.~(\ref{eq:donor-full}) and~(\ref{eq:inter-full}) we do not include explicitly
the pinning point for the plane-wave part of the
Bloch functions at the the donor and interface potentials.
Effects of the
interference induced by the different pinning points are
discussed in Ref.~[\onlinecite{calderonPRB08}]. We solve for the lowest
states of the full potential (donor, electric field and interface) by
obtaining the spectrum of $H$ in the combined basis of the lowest $D$ and $I$
envelopes as determined variationally, each multiplied by the respective
Bloch functions. This defines the Hilbert space for our  model calculation.
States with different Bloch indices $\mu\ne\nu$ are not coupled unless there
is a non-zero contribution from $H_{\rm vo}$ (see Appendix A).
Within the
Hilbert space defined here, the Hamiltonian is represented by a $12 \times 12$ matrix,
written formally as four $6\times 6$ blocks

\begin{equation}
H=
\left[{\begin{array}{cc}
 H_{DD} & H_{ID}\\
 H_{DI} & H_{II}
\end{array}}\right]
\label{eq:matrix}
\end{equation}
The equation giving the spectrum takes into account the non-orthogonality of
our basis, \textit{i.e.}, we solve for $H\Psi_{i}=E_{i}S\Psi_{i}$, where  $S$ is the
12$\times$12 overlap matrix so that $S_{DD}^{\mu,\nu} =
S_{II}^{\mu,\nu}=\delta_{\mu\nu}$ and $S_{DI}^{\mu,\nu} =
S_{ID}^{\mu,\nu}=\delta_{\mu\nu}\left<{F_{D}^\mu \mid
F_{I}^\mu}\right>$. Here,  $\left<{F_{D}^\mu \mid F_{I}^\mu}\right>$ is the
overlap between the interface and the donor envelope functions, an exponentially
decreasing function of $d$. A general state is
written as:
\begin{equation}
\Psi=\sum_{\mu= \pm x,\pm y,\pm z}\sum_{L=I,D}C_L^{\mu}F_{L}^{\mu}(\mathbf{r})\phi_{\mu}(\mathbf{r}) \, .
\label{eq:full}
\end{equation}
The complex coefficients $\{C_L^{\mu}\}$ give the contributions to the state
$\Psi$ of each valley $\mu$ at the interface or donor $(L=I,D)$. Also, they
define the symmetry of the state, and are referred to here as ``valley
coefficients''. The normalization condition reads
\begin{equation}
\sum_{\mu,\nu,L,L'} C_L^{\mu*} S_{LL'}^{\mu,\nu} C_{L'}^\nu = \sum_L |C_L^\mu|^2+{\rm cross~terms} = 1,
\label{eq:norm}
\end{equation}
where the cross terms are zero for $L=L'$ and/or for $\mu\ne\nu$.
For $d=4a^*$ the cross terms due to the finite overlap between donor and interface envelopes are less than $5\%$
in the electric field range of interest. The overlap $S_{ID}$ is negligible for $d=5a^*$ and larger.

In Fig.~\ref{Esquema}, we show the valley coefficients of the states at an
isolated donor (just the $\{C_D^\mu\}$ are given since $C_I^\mu =
0$ for all $\mu$)  and at an interface under a perpendicular electric field
(just the $\{C_I^\mu\}$ are given since $C_D^\mu = 0$ for all
$\mu$). The valley coefficients are perturbed when the impurity is located at
a distance $d$ from the interface, and hybrid states may be formed, \cite{martins04,calderonPRL06}  where both $D$ and $I$ coefficients contribute. The different initial (for low $F$) and final (for strong enough $F$) states are symmetry-compatible when their valley compositions are not orthogonal [see Fig.\ref{Esquema} (main panel)].

\section{Results}
\label{sec:results}

\begin{figure}
\includegraphics[clip,width=0.45 \textwidth]{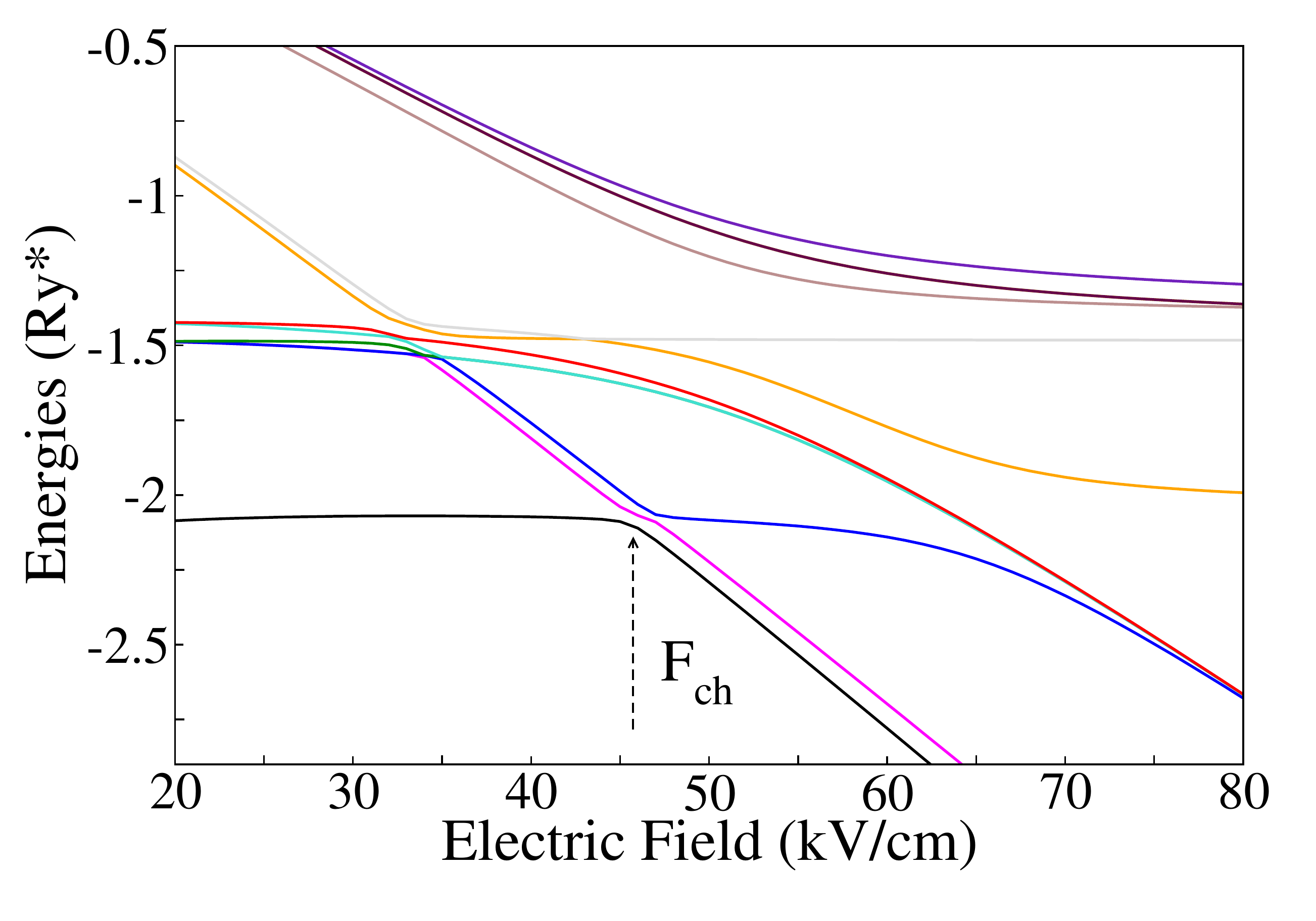}
\caption{
(Color online) Full spectrum of eigenvalues for $d=4a^*$  in a wide range of electric fields.
For small values of the electric field,  the lowest six eigenvalues correspond to
the donor states and the highest six to interface states. We use here $V_{\rm vo}^I=|V_{\rm vo}^I| e^{i \pi/3}$ with $|V_{\rm vo}^I|=\lambda F$ and $ \lambda =1.36$ \AA.
}
\label{fig:eigenvaluesd4}
\end{figure}

Previous theoretical and experimental studies
\cite{martins04,calderonPRL06,rahman2009,lansbergen-NatPhys} identified and analyzed a
characteristic field  at which the electron ground state crosses over from
being bound mainly around the donor (donor-like) into being bound mostly near
the interface (interface-like). This field, $F_{\rm ch}$ (see Fig.~\ref{fig:eigenvaluesd4}), decreases with  the distance $d$ from donor to interface, while the tunneling time increases
exponentially with $d$.  These properties are consistent with our 6-valley
model results. We explore here the valley degrees of freedom, mainly close to
the characteristic field, and analyze how the applied field changes the
valley composition of the main electronic states involved in the
donor-interface shuttling.

The electric-field dependence of the complete spectrum of the Hamiltonian in
Eq.~(\ref{eq:matrix}), for $d=4 a^*$, is shown in
Fig.~\ref{fig:eigenvaluesd4}. The overall level structure here is  similar to
the one presented in Refs.~[\onlinecite{debernardi06,rahman2009}]. At small fields, $F
\lesssim 32$ kV/cm, the six lowest eigenstates correspond to donor-bound
states. The degeneracies of the $T_2$ triplet and the $E$ doublet are lifted
(with a very small splitting) due to the perturbation produced by the interface. At
large fields $F \gtrsim 70$ kV/cm, the six lowest eigenstates become
interface-bound states. The two lowest eigenstates, which are combinations of
the $z$ and $-z$ valleys, are split by $2 |V_{\rm vo}^I|$. Different eigenvalues cross over each
other as $F$ changes. The levels cross or anticross depending on their relative symmetry. The general scheme shown in
Fig.~\ref{fig:eigenvaluesd4} is qualitatively similar for different values of
the distance $d$. The size of the gap at the anti-crossings is related to the shuttling time and decreases as $d$ increases.\cite{calderonPRL06}

\begin{figure}
\includegraphics[clip,width=0.5 \textwidth]{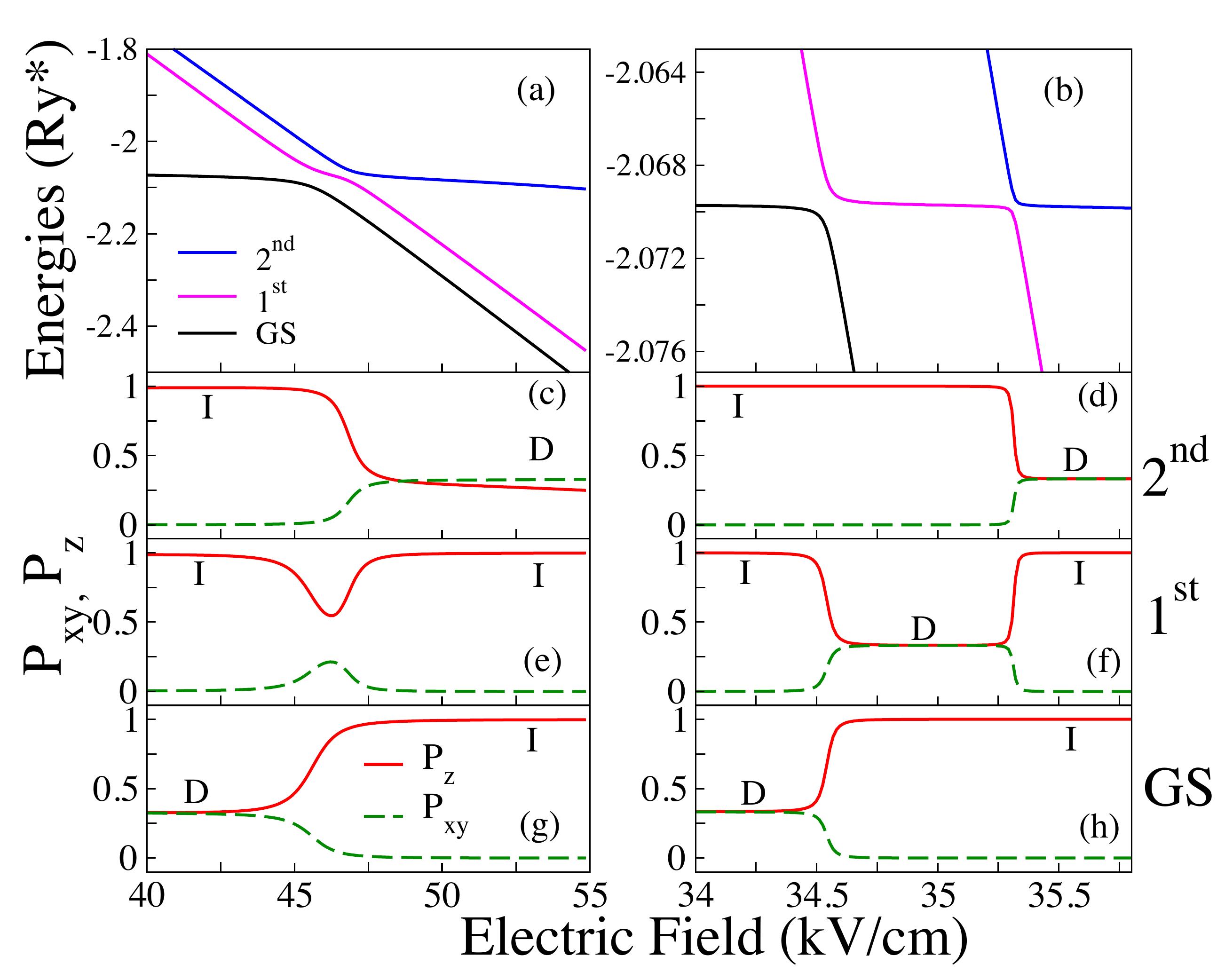}
\caption{(Color online) Evolution of the valley population for the three lowest eigenvalues around the characteristic field (see Fig.~\ref{fig:eigenvaluesd4}) for two different distances (a) $d=4a^*$ and (b) $d=5a^*$. We use  $V_{\rm vo}^I=|V_{\rm vo}^I| e^{i \pi/3}$ with $|V_{\rm vo}^I|=\lambda F$ and $\lambda =1.36$ \AA. The top panels reproduce the  eigenvalues involved in the lowest energy anti-crossing, which is in fact a sequence of two anti-crossings.  All other panels show the valley population in the second excited state, the first excited state, and the ground state (GS) in different lines.  The red (solid) curves correspond to the weight of the $\pm z$ valleys or longitudinal weight (at donor and interface), and the green (dashed) curves are the weight of the $\pm x$ and $\pm y$ valleys or transversal weight (at donor and interface). Labels D or I refer to donor-like or interface-like states in terms of real space location. Here D is a combination of the 6 valleys and I involves the z and -z valleys.
}
\label{fig:eigenvectorsd4d5}
\end{figure}

We follow now on the evolution of the valley contributions of the
three lowest eigenstates, which,  via mutual energies crossing over,
``become'' or contribute to the ground state for some range of field values (see Fig. ~\ref{fig:eigenvectorsd4d5}).
Because of time-reversal symmetry, valleys $\mu$ and $-\mu$ contribute
equally to any given eigenstate, so we may quantify the valley contributions by valley populations of each direction $\nu = x,~y,~z$, defined as
\begin{equation}
P_\nu = \sum_{\mu=\pm  \nu}\sum_{L=I,D} |C_L^\mu|^2 \, .
\label{eq:Pz}
\end{equation}
Due to the cylindrical symmetry of the system,  $P_z$ is in general
different from $P_x$ and $P_y$, while the two latter are equivalent. This
means that states with exchanged coefficients $C_{L=I,D}^{\pm
x}\leftrightarrow C_{L=I,D}^{\pm y}$ give the same expectation value for the
energy, since $H$ is invariant under ${\pm x \leftrightarrow \pm y}$. So $P_x$ and $P_y$  are presumably equal. It may
occur for a particular state that the weights $P_x$ and $P_y$ obtained
numerically differ: in this case a degenerate state is always found with
$P'_x=P_y$ and $P'_y=P_x$, as expected. Therefore, differences in $P_x$ and $P_y$ are not
physically meaningful and  we present our results in terms of $P_z$ and
\begin{equation}
P_{xy} = (P_x+P_y)/2 = {1\over 2}\sum_{\mu=\pm x, \pm y }\sum_{L=I,D} |C_L^\mu|^2 \,.
\label{eq:Pxy}
\end{equation}
In this definition we  do not take  the cross terms from Eq.~(\ref{eq:norm}) into account, so  normalization gives $2P_{xy} + P_z \approx 1$, allowing $P_z$ and $P_{xy}$ to be directly compared to each other, giving the relative weight of the $z$ and the average $x$ and $y$ populations. The lowest interface state at large $F$ only involves $z$ and $-z$ valleys, therefore $P_{xy}=0$ and $P_z=1$. On the other hand, the lowest donor state involves a symmetric combination of all valleys leading to $P_{xy}=P_z$. Hybrid states correspond to intermediate values of $P_{xy}$ and $P_z$: $0<P_{xy}<1/3$ and $1/3<P_z<1$.

In Fig.~\ref{fig:eigenvectorsd4d5} (upper frames) the spectrum is presented for a reduced range of energy and fields around the characteristic field for (a) $d=4 a^*$ ($F_{\rm ch}\sim 46$ kV/cm) and (b) $d=5 a^*$ ($F_{\rm ch}\sim 35$ kV/cm). The three rows of frames below give the corresponding valley populations for the $2^{\rm nd}$ excited, $1^{\rm st}$ excited, and ground (GS) states, respectively.  Here we take a complex $V_{\rm vo}^I=|V_{\rm vo}^I| e^{i \pi/3}$.  From Fig.~\ref{fig:eigenvectorsd4d5}  one can clearly observe that the anti-crossing at $F_{\rm ch}$ in fact involves two anti-crossings: one between GS and $1^{\rm st}$ excited, and another one between $1^{\rm st}$ and $2^{\rm nd}$ excited states.   Well below $F_{\rm ch}$, the GS is donor-like with $P_{xy}=P_z$. Above the two anti-crossings the GS and $1^{\rm st}$ excited are interface-like states, only involving the $z$ and $-z$ valleys, thus $P_{xy}=0$. 

We note that the ground state and the $2^{nd}$ excited state swap their valley compositions for fields below and above the crossover region. In fact, comparison of different frames in Fig.~\ref{fig:eigenvectorsd4d5} show that the low-field D-like composition in (g) "moves" to the high-field behavior in (c), while the low-field I-like composition in (c) is found in the high-field behavior in (g). Along the crossover region the compositions change smoothly or abruptly (according to $d$) among the limiting behaviors. The same applies to frames (d) and (h) at a more distant donor position from the interface. The intermediate state, shown in (e) and (f), preserves the compositions at  low and high fields, while it is clear that this state hybridizes with both ground and $2^{nd}$ excited states along the crossover region. In summary, the net effect of the field on the three lowest eigenfunctions far from $F_{ch}$ is to cross the ground and $2^{nd}$ excited states, while the intermediate $1^{st}$ excited state is not affected overall, although all three states mix at the crossover range. An extreme example of the $1^{st}$ excited state not being affected by the crossover is discussed below, in the context of Fig.\ref{fig:S&N}.

Comparison between (a) and (b) in Fig.~\ref{fig:eigenvectorsd4d5} illustrates the well known effect of increasing $d$, already discussed in previous publications, \cite{calderonPRL06,calderon-longPRB07} namely reducing anti-crossing gaps and sharpening transition lines.  Between the two anti-crossings, the $1^{\rm st}$ excited state is donor-like for $d=5 a^*$ in Fig. ~\ref{fig:eigenvectorsd4d5}(f), however for $d=4 a^*$ in (e) the $1^{\rm st}$ excited state is an interface-donor hybrid due to the larger donor-interface overlap. We note that all the gaps obtained for $d=5a^*$ are extremely small compared to the relevant energy scales here, so for most practical purposes the behavior would be equivalent to level crossing.

\begin{figure}
\includegraphics[clip,width=0.5 \textwidth]{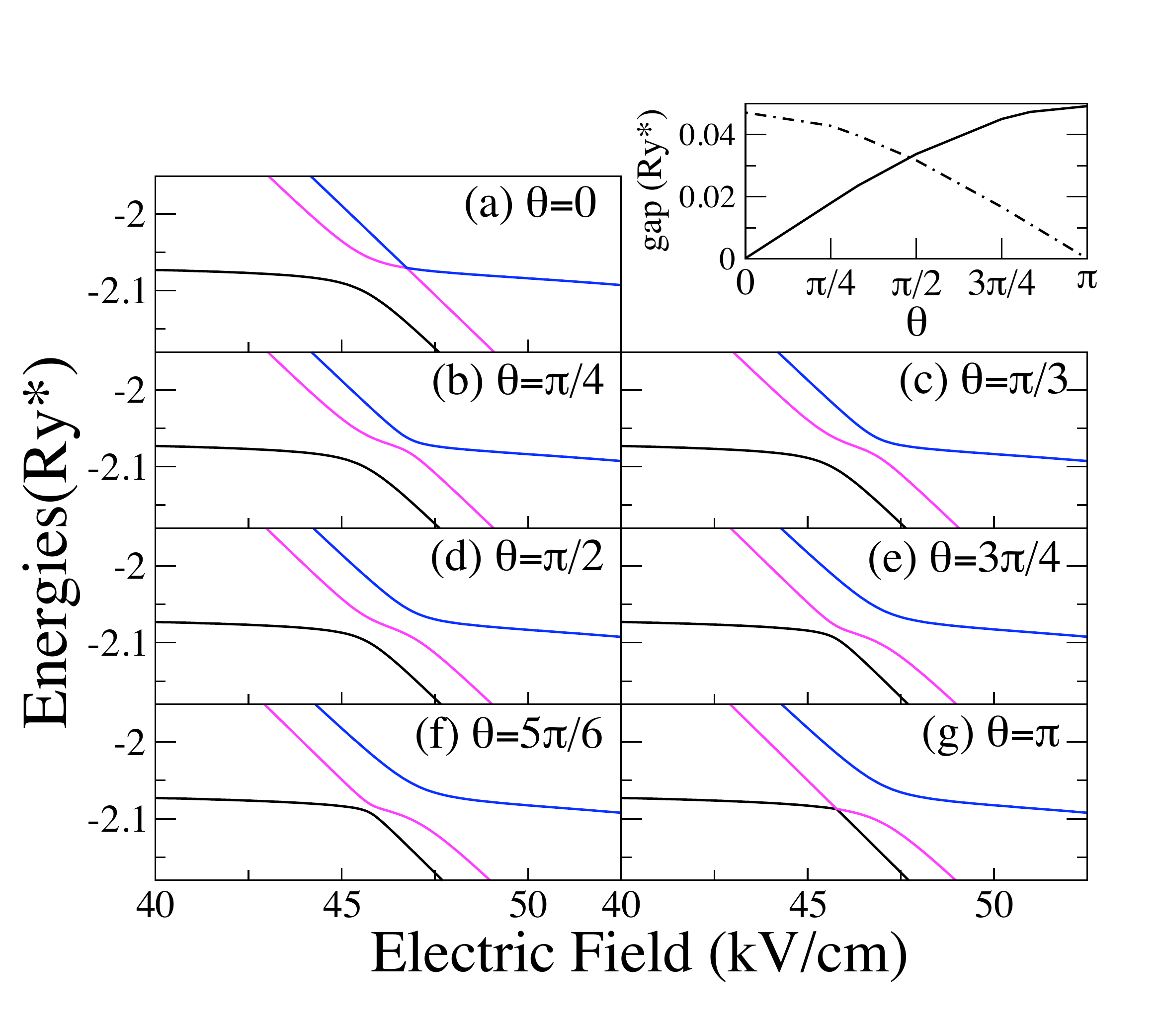}
\caption{(Color online) Three lowest eigenvalues around the anti-crossings region, corresponding to $d=4a^*$ for different values of the phase $\theta$ of the valley-orbit coupling at the interface ($V_{\rm vo}^I=|V_{\rm vo}^I| e^{i \theta}$).  $|V_{\rm vo}^I|=\lambda F$ and $\lambda=1.36$ \AA. The extra panel on the right top corner shows the values of the two anti-crossing gaps versus $\theta$. The dashed line represents the gap for the first anti-crossing between the GS and the $1^{\rm st}$ excited state while the solid line is the gap between the $1^{\rm st}$ and $2^{\rm nd}$ excited states. Note that for $\theta=0$ and $\pi$,  $V_{\rm vo}^I$ is a real quantity and one of the anti-crossings has zero gap (namely, it is actually a 2-level crossing). The two gaps are equal for $\theta=\pi/2$ which corresponds to a pure imaginary $V_{\rm vo}^I$. Our results are obviously invariant for  $\theta \leftrightarrow -\theta$.}
\label{fig:d4-vs-theta}
\end{figure}

\begin{figure}
\includegraphics[clip,width=0.3 \textwidth]{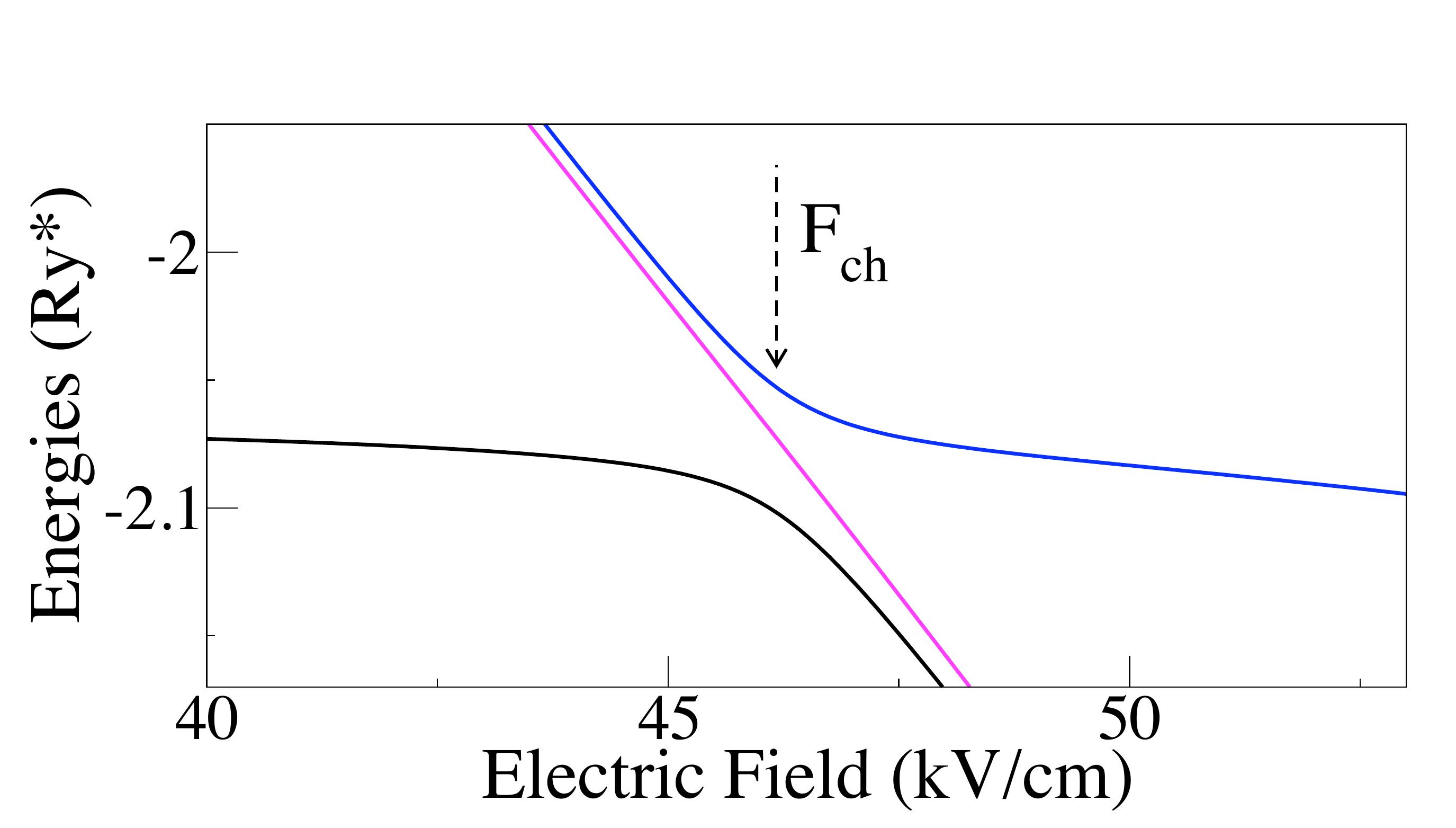}
\caption{(Color online) Three lowest eigenvalues for $d=4a^*$ close to the characteristic field with the valley orbit coupling at the interface $|V_{\rm vo}^I|=\lambda F$ with $\lambda=0.215$~\AA~  as calculated by Sham and Nakayama. \cite{sham79} We use $\theta=\pi/3$ as the phase of $V_{\rm vo}^I$  [compare with Fig.~\ref{fig:d4-vs-theta}(c)].
}
\label{fig:S&N}
\end{figure}

The phase $\theta$ on $V_{\rm vo}^I=|V_{\rm vo}^I| e^{i \theta}$, not as extensively considered so far, is analyzed in Fig.~\ref{fig:d4-vs-theta} for $d=4a^*$. The complex phase of the valley-orbit coupling at the interface affects the symmetries of the eigenstates leading to different gaps at the two anti-crossings around $F_{\rm ch}$. Figs.~\ref{fig:d4-vs-theta}(a) and (g) show the limiting cases of $\theta=0$ and $\theta=\pi$, which correspond to a real $V_{\rm vo}^I$, lead to a zero-gap (crossing) involving the symmetric donor-like eigenstate and the antisymmetric interface state, with all $C_L^\mu =0$, except $C_I^z=-C_I^{-z}=1/\sqrt{2}$. For a general $\theta$, the two lowest interface states always have a symmetric component which "repel" the symmetric donor-like level. The size of the gap at the anti-crossing increases as the weight of the symmetric part of the interface-like state becomes larger. The two gaps become equal for $\theta=\pi/2$ which corresponds to a purely imaginary  $V_{\rm vo}^I$. In summary, for a fixed $|V_{\rm vo}^I|$ and  calling $g_L$ and $g_R$ the gaps to the left and to the right in energy, we get $g_L>g_R$ for $0<\theta<\pi/2$ and $g_R>g_L$ for $\pi/2<\theta<\pi$. In particular $g_R =0$ ($g_L=0$) for $\theta = 0$ $(\pi)$ and $g_L=g_R$ for $\theta = \pi/2$. The largest gap observed for $d=4 a^*$ is $\sim 1$ meV, of the same order than the interface valley-orbit splitting considered. In contrast,  for $d=5a^*$ the gap is $\sim 0.035$ meV, almost two orders of magnitude smaller.

The results presented so far correspond to a relatively large value of the valley-orbit coupling at the interface, with $\lambda=1.36$ \AA~ as estimated in Ref.~[\onlinecite{chutia2008}]. For this case and $d=4a^*$, the gap at anti-crossing is slightly smaller than the value of the valley-orbit splitting ($2 |V_{\rm vo}^I|$).
Previous calculations by Sham and Nakayama \cite{sham79} lead to a smaller $\lambda= 0.215$\AA. In Fig.~\ref{fig:S&N} the three lowest eigenvalues close to the characteristic electric field for $d=4a^*$ for the $V_{\rm vo}^I$ as calculated by Sham and Nakayama are shown. Here, the gap at anti-crossing   (at $F=F_{\rm ch}$) is about three times larger than the
value of the valley-orbit splitting at the  interface for $d=4a^*$. Due to the relatively smaller value of the valley-orbit splitting, the interface states are much closer: The two anti-crossings at $F_{\rm ch}$ seem to merge into a single one and the $1^{\rm st}$ excited state is always an interface state. This is in contrast with results for the larger value of $V_{\rm vo}^I$ where the $1^{\rm st}$ excited state is hybridized or donor-like between the two anti-crossings [see Fig.\ref{fig:eigenvectorsd4d5}(e) and (f)].
Another distinct feature of the small $ |V_{\rm vo}^I|$ limit, illustrated in Fig. \ref{fig:S&N}, [compare with Fig. \ref{fig:d4-vs-theta}(c)],
 is the way the three levels separate around  $F_{\rm ch}$, with level repulsion among the outer ones,
and no deviation of the middle state from the linear path: it does not couple to the others. Fig. \ref{fig:S&N} is very similar to the inset of Fig. 2 in Ref. [15], where the model does not include a barrier, thus corresponding to the $ V_{\rm vo}^I = 0 $ limit. We may infer from these results that different ionization regimes for doped Si may arise as a function of d (inversely related to the gap at anticrossing) and the valley-orbit coupling at
interface (related to the electric field, to the height, and to the quality of the interface barrier).

\section{Discussion}
\label{sec:discussion}

We have analyzed in full depth the valley contributions to the three lowest energy levels of a single
electron in the donor-in-Si-near-a-barrier system under an applied external field, focusing in the field range around the crossover between donor-like and interface-like character of the ground state. Strong hybridization occurs in the vicinity of the crossover, which we identify as a sequence of two anti-crossings  (see Fig.~\ref{fig:eigenvectorsd4d5}). A result to keep in mind in practical applications is the strong dependence of the levels crossings and anti-crossings on the phase $\theta$ of the valley-orbit coupling at the interface, $V_{\rm vo}^I = |V_{\rm vo}^I| e^{i\theta}$. This is to be expected since the phases affect directly the symmetry of the states. A trivial example is the case $\theta = \pi$ (0), where the GS is a symmetric (anti-symmetric) combination of the $z$ and $-z$ valleys. It is not a straightforward task  to predict or control the phase $\theta$, which should  vary with the barrier material,\cite{Andre2011}  interface roughness due to steps, interdiffusion, etc, and other sample properties.

In practice, the results on the double anti-crossings shown in Fig.~\ref{fig:d4-vs-theta} may play an important role
in applications involving the donor/barrier system. For instance, in the limit
of $\theta\rightarrow 0$, the ground state is well separated from the excited states, suggesting
the possibility of adiabatically shuttling the electron from the donor site to the barrier interface. This could be a suitable scenario for a spin qubit for which one needs the valley splitting to be larger than the Zeeman splitting. An intrinsically large $|V_{\rm vo}^I|$, like the one considered in Figs.~\ref{fig:eigenvaluesd4}, \ref{fig:eigenvectorsd4d5}, and \ref{fig:d4-vs-theta}, is also needed, as well as a small $d$ ($\lesssim 4a^*$) to guarantee a sufficiently large gap at anticrossing.

On the other hand, following the adiabatic
theorem, the donor-interface shuttling would be prohibitively slow for $\theta\rightarrow \pi$ due to the crossing between the ground state and the first excited level. Another phase-related effect occurs due to the variable pinning point in the Bloch functions $\phi_\mu(\mathbf{r})$ [\onlinecite{koiller04,calderonPRB08}], which produces a periodic dependence of
the valley-orbit splitting on $d$ leading to the closing of the gap at anticrossings at some particular values of $d$  [\onlinecite{calderonPRB08}].

If instead of the spin we want to use the valley degree of freedom to define qubits, a controllable valley-orbit coupling is required.\cite{dimi2012}
This is achieved if the energy separation of two states with the
same envelope function but different valley compositions (therefore, different oscillations in the atomic scale)
varies significantly with the external applied fields. This situation can be attained above $F_{\rm ch}$ when the two lowest eigenstates are mainly interface states involving different combinations of the $z$ and $-z$ valleys.  A strong dependence of the splitting with electric field can be found close to $F_{\rm ch}$ in some cases. For instance, in Fig.~\ref{fig:d4-vs-theta}(g) (corresponding to $\theta=\pi$) the closing of the gap between the two lowest eigenvalues produces a fast decrease of the splitting as the electric field is lowered from $\sim 48$ to $\sim 46$ kV/cm.  Also, for smaller values of $V_{\rm vo}^I$, as exemplified in Fig.~\ref{fig:S&N}, the interface-donor hybridization leads to a relatively  large gap compared to $V_{\rm vo}^I$ and consequently the splitting increases fast from $\sim 48$ to $\sim 46$ kV/cm.  However, the variation of the level splitting on the electric field is stronger when the donor-interface hybridization is large. This hybridization mixes valleys in the different directions, potentially producing decoherence in the valley sector.

On the other extreme, $\theta=0$,  Fig.~\ref{fig:d4-vs-theta}(a) reveals that the upper valley composition crosses the second excited
state: As a consequence, the valley information would be lost.

In summary, we show that both the modulus and the phase of $V_{\rm vo}^I$ affect the quantum behavior of donor electrons in Si near an interface.  The value of $\theta$ is hard to predict, and its calculation
probably requires knowledge of the atomistic distribution at the interface and a complete electronic structure description of the combined Si/structured interface/barrier system. It is not clear whether a direct  experimental measure is possible. We propose that measurement of the separate gaps may provide an estimate of $\theta$, as shown in the inset of Fig.~\ref{fig:d4-vs-theta}. Finally, the trends and discussions regarding Fig.\ref{fig:eigenvectorsd4d5} should bring new and valuable insight towards controlled valley manipulations.

A.B. and M.J.C. were supported by FIS2009-08744 (MINECO, Spain). AS and BK's  work is part of  the Brazilian National Institute for Science and Technology on Quantum Information. AS and BK acknowledge partial support from FAPERJ, CNPq and CAPES
\\

\appendix

\section{Valley-orbit term ($H_{\rm vo}$)}
 The last term in Eq.~(\ref{eq:hamil}), with $\Delta_\perp=-\Delta_c=-2.16$ meV, $\Delta_\parallel=-\Delta_c(1+\delta_c) = -1.51$ meV,  and  $|V_{\rm vo}^I|=\lambda F$ with $\lambda=1.36$ \AA.
This is a 12x12 matrix in the basis set of the six valleys at the donor and the six valleys at the interface.  According to the notation in Eq.~(\ref{eq:matrix}):
\begin{equation}
[H_{\rm vo}]_{DD}=
\left(\begin{array}{cccccc}
0 & \Delta_\parallel &\Delta_\perp &\Delta_\perp &\Delta_\perp & \Delta_\perp\\
\Delta_\parallel & 0  &\Delta_\perp  &\Delta_\perp &\Delta_\perp &\Delta_\perp \\
\Delta_\perp &\Delta_\perp& 0 &  \Delta_\parallel &\Delta_\perp&\Delta_\perp \\
\Delta_\perp & \Delta_\perp & \Delta_\parallel & 0 &\Delta_\perp &\Delta_\perp \\
\Delta_\perp  & \Delta_\perp  &\Delta_\perp  &\Delta_\perp  & 0 & \Delta_\parallel \\
\Delta_\perp & \Delta_\perp &\Delta_\perp & \Delta_\perp & \Delta_\parallel &0\\
\end{array} \right)
\end{equation}
\bigskip
\begin{equation}
[H_{\rm vo}]_{II}=
\left(\begin{array}{cccccc}
0 & 0 & 0 & 0 & 0 & 0 \\
0 & 0 & 0 & 0 & 0 & 0  \\
0 & 0 & 0 & 0 & 0 & 0 0 \\
0& 0& 0 & 0 & 0 & |V_{\rm vo}^{I}|e^{i\theta} \\
0 & 0 & 0 & 0 &  |V_{\rm vo}^{I}|e^{-i\theta} & 0
\end{array} \right)
\end{equation}

and $[H_{\rm vo}]_{ID}=[H_{\rm vo}]_{DI}=0$.

\end{document}